# Nanocrystalline structure and strain in magnesium under extreme dynamic compression

D.A. Komkova[1*], A.Yu. Volkov[1], and E.F. Talantsev[1,2]

[1]M.N. Miheev Institute of Metal Physics, Ural Branch, Russian Academy of Sciences, 18, S. Kovalevskoy St., Ekaterinburg, 620108, Russia

[2]NANOTECH Centre, Ural Federal University, 19 Mira St., Ekaterinburg 620002, Russia

*Corresponding author: komkova_d@imp.uran.ru

**Abstract.** The study of materials behavior under extreme conditions is fundamental to science and modern technology. Fast ramp compression is a unique method for exploring materials behavior and phase transformations under extreme conditions. One unexplored feature of this method is the nanoscale structure of the material under dynamic compression. This leaves a gap in understanding the details of phase transformations under fast ramp compression. Here, we made a first step in the exploration by applying the Williamson-Hall (WH) analysis to X-ray diffraction data (XRD) measured in magnesium subjected to fast ramp compression at four pressures. We found that at $P = 309\,GPa$ magnesium in *bcc*-like phase has an average crystalline size $D = (2.2 \pm 0.7)\,nm$ and microstrain $\varepsilon = (-0.011 \pm 0.007)$. At $P = 409\,GPa$, magnesium demonstrates $D = (4.5 \pm 3)\,nm$ with $\varepsilon = (-0.003 \pm 0.007)$. At $P = 563\,GPa$, *Fmmm*-magnesium has crystalline size $D = (2.6 \pm 0.5)\,nm$ with microstrain $\varepsilon = (-0.004 \pm 0.004)$. At $P = 959\,GPa$, we revealed that *sh*-magnesium exhibits average size of $D > 12\,nm$ and relatively high value of microstrain $\varepsilon = (0.011 \pm 0.002)$. In the result, we report the first microstructural evolution insights of magnesium under fast ramp compression.

**Keywords:** magnesium, fast ramp compression, Williamson-Hall analysis, nanostructure, microstrain.

# Nanocrystalline structure and strain in magnesium under extreme dynamic compression

## 1. Introduction

Magnesium, the lightest structural metal, has an exceptional strength-to-weight ratio, corrosion resistance, and recyclability that make the metal of significant fundamental and practical interest [1–3]. Under ambient conditions, it demonstrates anisotropic behavior and poor ductility because of a hexagonal close-packed (*hcp*) structure [4]. To improve Mg formability, research teams have been focused on refining Mg microstructure through severe plastic deformation (SPD) methods. Such well-known methods as equal-channel angular extrusion [5], high pressure torsion [6], rolling [7], or non-conventional lateral extrusion [8,9] allow for achieving ultrafine-grained and nanocrystalline structures with high defect densities and grain boundaries, which can significantly enhance mechanical properties.

However, the microstructure produced by SPD is typically limited to grain sizes above 100-300 nm [6,10], and these are achieved under quasi-static to moderate strain-rate conditions. At the same time, the evolution of the microstructure of magnesium at extreme pressures above 100 GPa remains virtually unexplored. Such conditions, far beyond the SPD regime, as a rule, generate profound material responses [11–14], including phase transitions and outstanding hardening. For example, Mg is known to undergo a series of complex transitions with increasing pressure from body centered cubic (*bcc*) phase at approximately 50 GPa [15,16] to simple cubic (*sc*) phase at pressures exceeding 660 GPa [17,18]. In the recent past, Gorman *et al.* [19] experimentally revealed the sequence of Mg phase transition from *bcc* phase to *sc* phase at pressures from 309 to 1317 GPa under fast ramp compression. Moreover, under strain rates over $10^6$ s$^{-1}$, the pure metals like copper [20], titanium [21], lead [14] are shown to exhibit dramatically increased hardening due to ballistic transport of dislocations, limited by phonon drag [21].

Though X-ray diffraction (XRD) during such experiments can identify phases and lattice parameters, the direct microscopic observation of the atomic structure formed under ramp compression (e.g., grain size, microstrain, dislocation density) is impossible and remains unexplored field to date. Meanwhile, plastic deformation, phase transitions, and relaxation processes involve grain refinement and the generation and redistribution of internal stresses. Lack of knowledge of these characteristics in materials under fast ramp compression leads to an inability to understand and predict the behavior of materials under extreme conditions.

One of the possible ways to explore this *terra incognita* was recently proposed [22], where it was suggested that the Williamson-Hall (WH) analysis [23] can be applied to experimental XRD data recorded during the ramp compression of materials. Because the WH analysis is the most basic, robust, and widely used analysis to determine the crystalline size and strain from XRD data, it is important to further explore its applicability for the analysis of XRD data measured in fast ramp compressed materials.

In the present work, we extended the application of the of the WH analysis to fast ramp compressed metallic elements by applying the method to XRD data registered during the dynamic compression of elemental magnesium at 309 GPa, 409 GPa, 563 GPa and 959 GPa [19].

## 2. Experimental data sources and data analysis method

XRD data for fast ramp-compressed Mg, reported by Gorman *et al.* [19], were analyzed in this study. For this purpose, we digitized the experimental curves for Mg under 309 GPa, 409 GPa, 563 GPa, and 959 GPa presented in Fig. 2b [19].

Approximations and fittings were performed with the Origin software (OriginPro, Version 2021, OriginLab Corporation, Northampton, MA, USA). A profile of each diffraction peak was fitted to the Gauss peak function:

$$(2\theta) = I_{background} + \frac{A}{w\sqrt{\frac{\pi}{2}}} e^{\left(-2\left(\frac{2\theta - 2\theta_c}{w}\right)^2\right)}, \tag{1}$$

where $I_{background}$, $2\theta$, $A$, and $w$ are free-fitting parameters. Fits for analyzed XRD peaks are shown in the Supplementary Materials Figs. S1 – S4.

The integral breadth, $\beta_{integral}(2\theta)$, of the diffraction peak was calculated by the following equation:

$$\beta_{integral} = w \cdot \sqrt{\frac{\pi}{2}}, \tag{2}$$

To extract the broadening caused by sample structural parameters, $\beta_s(2\theta)$, the standard equation was used [24,25]:

$$\beta_s(2\theta) = \sqrt{\beta_{integral}^2(2\theta) - \beta_{instr}^2(2\theta)}. \tag{3}$$

The XRD peak broadening due to sample effect contains contributions from both the average crystallite size, $D$, and the microstrain, $\varepsilon$. To separate and quantify these microstructural parameters for fast ramp compressed Mg at different pressures, we applied the WH analysis, and fitted calculated breadth, $\beta_s(2\theta)$, and corresponding diffraction angle, $2\theta$, of the XRD peaks to the equation [23]:

$$\beta_s(2\theta) = \frac{K \times \lambda_{XRD}}{D \times \cos\left(\frac{2\theta}{2}\right)} + 4 \times \varepsilon \times \tan\left(\frac{2\theta}{2}\right), \tag{4}$$

where $K$ is Scherrer constant as 0.9 [26], $\lambda_{XRD}$ is X-ray wavelength of 0.1209 nm [19].

All figures presenting fits with WH equation (Eq. 4) include the deduced parameters along with their standard errors and 95% confidence bands, calculated by standard procedures [27]. No additional statistical parameters for the fits were determined, as the limited number of raw experimental data sets prevented a more detailed analysis.

### 3. Instrumental broadening of the XRD diffractometer

It is known [28–30], the XRD peak breadth generally consists of physical broadening (sample effect), $\beta_s$, and instrumental broadening, $\beta_{instr}$. The fast ramp compression experimental results for elemental magnesium [19] were obtained at the National Ignition Facility [11,31].

Previously [22], we found that total instrumental broadening function, $\sigma_i(2\theta)$, presented by Rygg *et al.* for fast ramp compression [31] was significantly overestimated in comparison with experimental data in Fig. 1b [11], where the breadth of the diffraction peaks from Pt pinhole in XRD dataset of fast compressed gold turned out to be narrower than it was presented in Fig. 21 [31].

In the result, we analyzed the diffraction peaks from Pt pinhole (Fig. 1b in [11]) to determine parameters of the Caglioti-Paoletti-Ricci function [32] for the instrumental broadening function, $\beta_{instr}(2\theta)$, as follows:

$$\beta_{instr}(2\theta) = \sqrt{U \times \tan^2\left(\frac{2\theta}{2}\right) + V \times \tan\left(\frac{2\theta}{2}\right) + W}, \qquad (5)$$

where $2\theta$ is diffraction angle, $U = (4.9 \pm 0.3) \times 10^{-4}$ rad², $V = 0$ (fixed) rad², and $W = (3.9 \pm 0.1) \times 10^{-4}$ rad². Based on this, the resolution of the diffractometer, $D_{max}$, was found to be $D_{max} = 12\ nm$, establishing an upper limit for crystallite size measurement [22].

In the present study, we use the same function and parameters for $\beta_{instr}(2\theta)$ to determine $\beta_s(2\theta)$ by Equation 3, because experiments on fast ramp compressed magnesium were performed at the same experimental facilities as the ones in Refs. [11,31]. More detailed information of the instrumental broadening and resolution calculation is presented in [22].

### 4. Size and strain in fast ramp compressed elemental magnesium

#### *4.1. Size-strain analysis for magnesium at 309 GPa*

Phase states of magnesium at fast ramp compression at 309 GPa and 409 GPa are still under investigation. As Gorman *et al.* [19] reported, at 309 GPa and 409 GPa, the diffraction peaks corresponding to fast ramp compressed magnesium are inconsistent with the calculated *bcc* phase, and it is more likely that the phase is consistent with a *bcc*-like structure, details can be found in Ref. [19].

In this regard, it is worth noting that WH analysis does not require knowledge of the phase state of the material, since it relies on XRD peak breadth. Therefore, WH analysis can be applied to XRD data even when the phase state is not definitively established.

Our fits of the XRD peaks with the deduced parameters $w(2\theta, P = 309\ GPa)$ are shown in Fig. S1 (the fits were performed with Eq. 1). For a diffraction peak at $2\theta = 0.6955\ rad$ (or at $2\theta = 39.85°$, Fig. 2b and Extended Data Table 2 in Ref. [19]), we hypothesized that this wide peak is a convolution of two overlapping reflections, which were not resolved in the registered XRD data because of experimental challenges. The fit for the peak in the assumption of the two overlapped reflections is shown in Fig. S1, a

The fit of the deduced values $\beta_s(2\theta, P = 309\ GPa)$ with WH equation (Eq. 4) is presented in Fig. 1. Our analysis yielded that the crystallite size of the magnesium at this pressure is $D\ (P = 309\ GPa) = (2.2 \pm 0.7)\ nm$ with the microstarin of $\varepsilon\ (P = 309\ GPa) = (-0.011 \pm 0.007)$.

Negative values of microstrain probably requires a careful interpretation. According to [33], a value of $\varepsilon < 0$ suggests that strain is not a dominant source of broadening and may be considered negligible in comparison to size broadening, e.g., negative microstrain is reported mostly when the crystallite size is less than 20 nm [34]. In turn, other authors suppose that the negative microstrain is an indicator of compressive lattice distortions, and the magnitude increases as crystallite size decreases [35–37]. Moreover, the negative microstrain may also imply the presence of anisotropic effects, e.g., orientation-dependent microstrain or non-spherical crystallite shapes [38], which can cause the isotropic WH model to yield negative values or significant scatter, especially when only a few reflections are available. Here, we obtain a structure with a small grain size, $D \sim 2nm$, and negative microstrain of high magnitude, $|\varepsilon| \sim 0.011$, or $\sim 1.1\%$, under compression at $P = 309\ GPa$.

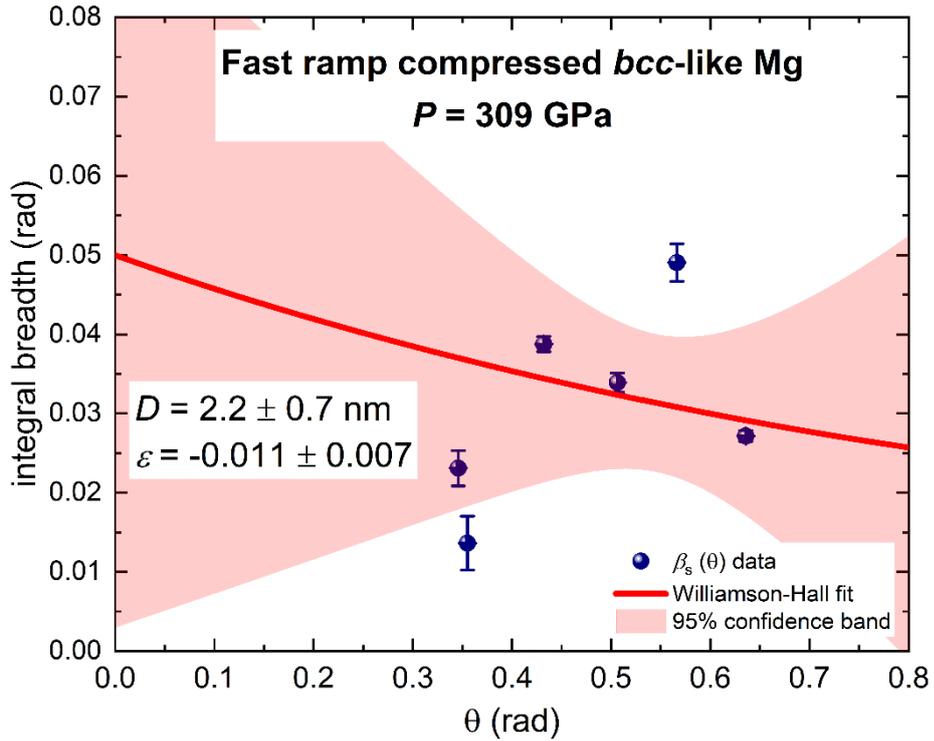

**Figure 1.** The fit of the $\beta s\ (\theta)$ data for fast ramp compressed magnesium at $P = 309\ GPa$ to the Williamson-Hall equation (Eq. 4) at $D$ and $\varepsilon$ are free-fitting parameters. Experimental XRD dataset was reported by Gorman *et al.* (Fig. 2b [19]). Deduced parameters are shown in the plot.

### *4.2. Size-strain analysis for magnesium at 409 GPa*

Fig. S2 presents the fits of the XRD peaks obtained from magnesium at $P = 409\ GPa$, along with the corresponding deduced parameters $w(2\theta, P = 409\ GPa)$. The fits were performed using Eq. 1. To the diffraction peak located at $2\theta = 0.7205\ rad$ (or at $2\theta = 41.28°$, Figure 2b and Extended Data Table 2 in Ref. [19]), the same assumption, as previously – that the observed broad profile results from two overlapping reflections unresolved due to experimental difficulties – was applied to the peak. The fit of the peak with two-peak model is shown in Fig. S2, a.

The fit of the deduced values $\beta_s(2\theta, P = 409\ GPa)$ with the WH equation (Eq. 4), where the crystallite size $D$ and microstrain $\varepsilon$ are free-fitting parameters, is provided in Fig. 2, a. The deduced microstrain $\varepsilon(P = 409\ GPa) = (-0.003 \pm 0.007)$ is four times lower than strain in the same phase at lower pressure, $\varepsilon(P = 309\ GPa) = (-0.011 \pm 0.007)$ with the same error as the deduced value.

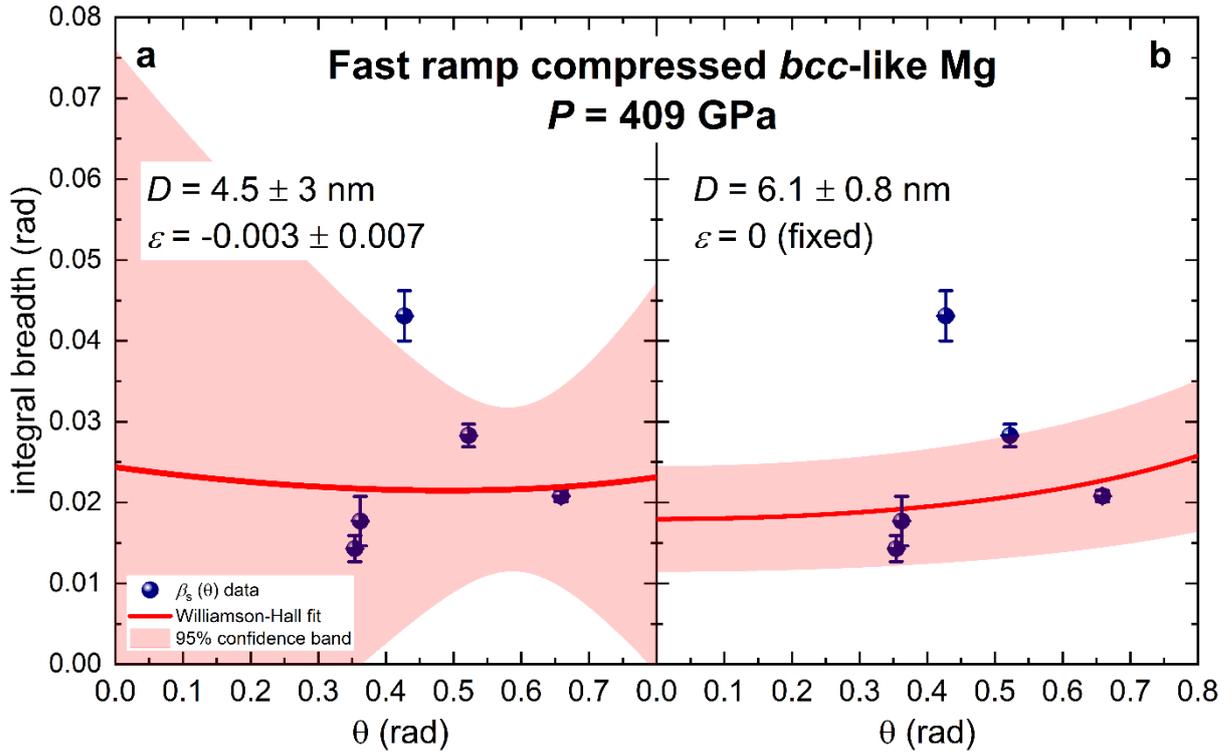

**Figure 2.** The fit of the $\beta s\,(\theta)$ data for fast ramp compressed magnesium at $P = 409\,GPa$ to the Williamson-Hall equation (Eq. 4). Experimental XRD dataset was reported by Gorman *et al.* (Fig. 2b [19]). Deduced parameters are shown in the plot. (a) $\beta s\,(\theta)$ fit where $D$ and $\varepsilon$ are free-fitting parameters. (b) $\beta_s\,(\theta)$ fit at $\varepsilon = 0$ (fixed).

Furthermore, while the crystallite size error is smaller than the value itself, or $D(P = 409\,GPa) = (4.5 \pm 3.0)nm$, microstrain error is seen to exceed the value, $\varepsilon(P = 409\,GPa) = (-0.003 \pm 0.007)$. Considering the above, it is reasonable to assume that the fit of $\beta_s(2\theta, P = 409\,GPa)$ can be performed with the WH model, where the microstrain is fixed to zero, $\varepsilon = 0$. The fit with this restriction (Fig. 2, b) allows determination of the crystalline size of magnesium under these ramp compression conditions with high accuracy, $D(P = 409\,GPa) = (6.1 \pm 0.8)\,nm$.

### 4.3. Size-strain analysis for magnesium at 563 GPa

As Gorman *et al.* [19] revealed, under ramp compression at 563 GPa, magnesium undergoes transformation into a new *Fmmm* phase, which represents orthorhombic distorted face-centered cubic (*fcc*) phase with a reduction in the *a* parameter and an increase in the *c* parameter.

Fig. S3 demonstrates the results of fitting the XRD peaks from magnesium at $P = 563\ GPa$ with Eq. 1. The deduced parameters $w(2\theta, P = 409\ GPa)$ are also listed. In case of $P = 563\ GPa$, the authors [19] considered the peak at $2\theta = 0.7367\ rad$ (or at $2\theta = 42.21°$, Figure 2b and Extended Data Table 2 in Ref. [19]) as two overlapping reflections corresponding to (002) and (111) peaks. Deconvolute peaks, however, have narrower breadth than the instrumental broadening. Thus, the deconvolution of the peak was not reliable, and the peak was fitted by single Gaussian peak function, as depicted in Fig. S3a.

Fig. 3 shows the results of fitting the $\beta_s(2\theta, P = 563\ GPa)$ parameters with WH equation (Eq. 4). The WH analysis with $D$ and $\varepsilon$ are free-fitting parameters (Fig. 3a) gives the crystallite size and microstrain deduced as $D(P = 563\ GPa) = (2.6 \pm 0.5)\ nm$ and $\varepsilon\ (P = 563\ GPa) = (-0.004 \pm 0.004)$, respectively. As seen, the microstrain $\varepsilon\ (P = 563\ GPa)$ is three times lower than that derived for magnesium at $P = 309\ GPa$, $\varepsilon(P = 309\ GPa) = (-0.011 \pm 0.007)$, and comparable to $\varepsilon(P = 409\ GPa) = (-0.003 \pm 0.007)$. Moreover, the error for $\varepsilon\ (P = 563\ GPa)$ and the value itself are equal. Therefore, it is appropriate to set $\varepsilon\ (P = 563\ GPa) = 0$, as it was done in case of $P = 409\ GPa$. Applying this condition to the fit of $\beta_s(2\theta, P = 563\ GPa)$ with Eq. 4 (Fig. 2b) provides a precise estimate of the crystallite size as $D(P = 563\ GPa) = (3.1 \pm 0.1)\ nm$.

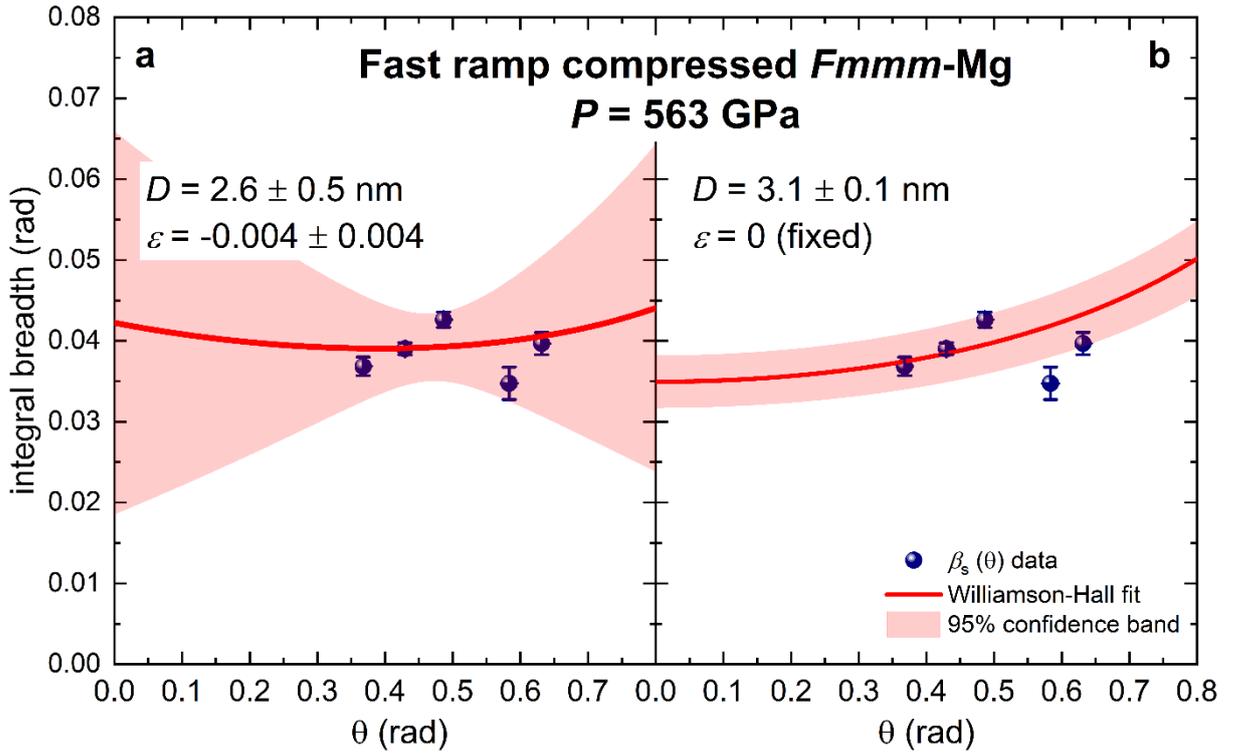

**Figure 3.** The fit of the $\beta s\ (\theta)$ data for fast ramp compressed magnesium at $P = 563\ GPa$ to the Williamson-Hall equation (Eq. 4). Experimental XRD dataset was reported by Gorman *et al.* (Fig. 2b [19]). Deduced parameters are shown in the plot. (a) $\beta s\ (\theta)$ fit where $D$ and $\varepsilon$ are free-fitting parameters. (b) $\beta_s(\theta)$ fit at $\varepsilon = 0$ (fixed).

### *4.4. Size-strain analysis for magnesium at 959 GPa*

The XRD peak fits with Eq. 1 and the deduced parameters $w(2\theta, P = 959\ GPa)$ are shown in Fig. S4. In the work of Gorman *et al.* [19], at $P = 959\ GPa$, two XRD peaks observed from the magnesium were consistent with a simple hexagonal (*sh*) structure, and one of the peaks was resolved into two overlapping reflections, namely the (001) peak at $2\theta = 0.7473\ rad$ (or $2\theta = 42.82°$) and (010) peak at $2\theta = 0.7669\ rad$ (or $2\theta = 43.94°$) (Fig. 2b and Extended Data Table 2 in Ref. [19]). Here, we followed the authors data, and the two-peak deconvolution with corresponding deduced parameters of $w(2\theta, P = 959\ GPa)$ listed below is depicted in Fig. S4a.

The fits of the $\beta_s(2\theta, P = 959\ GPa)$ with the WH equation (Eq. 4) can be seen in Fig. 4. $D$ and $\varepsilon$ are treated as free fitting parameters (Fig. 4), and the crystallite size $D(P = 959\ GPa)$ tends to infinity, indicating that the crystal size exceeds the instrumental resolution limit of $12\ nm$ [22]. Meanwhile, the microstrain in magnesium with *sh*-structure is $\varepsilon(P = 959\ GPa) = (0.011 \pm 0.002)$.

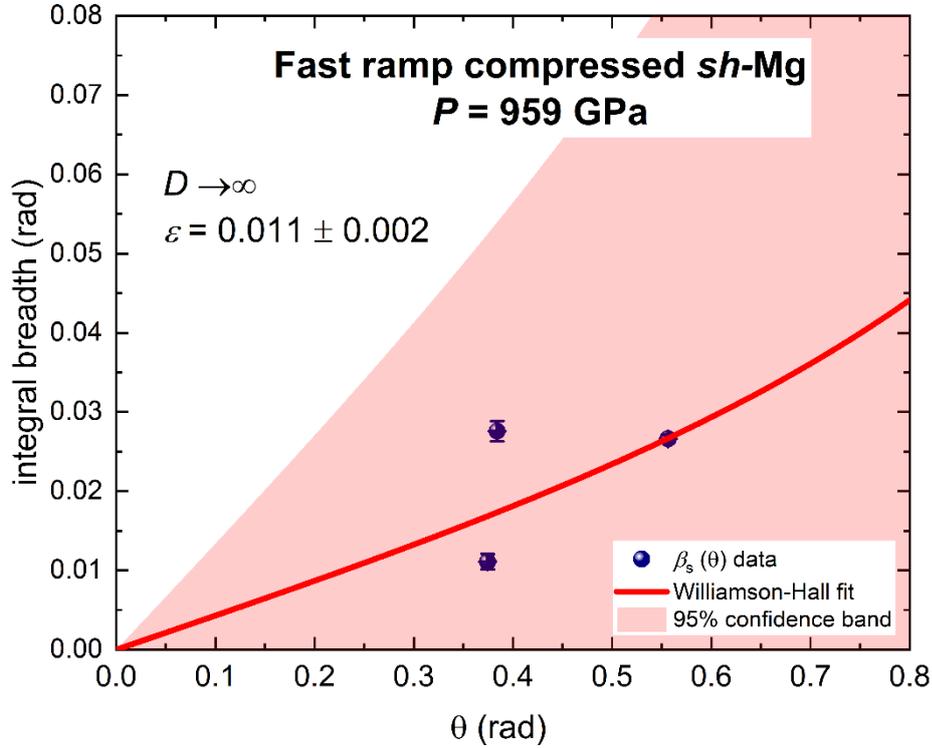

**Figure 4.** The fit of the $\beta s\ (\theta)$ data for fast ramp compressed magnesium at $P = 959\ GPa$ to the Williamson-Hall equation (Eq. 4). Experimental XRD dataset was reported by Gorman *et al.* (Fig. 2b [19]). Deduced parameters are shown in the plot.

The grain size, $D(P = 959\ GPa)$, of beyond $12\ nm$ and the positive high microstarin, $\varepsilon(P = 959\ GPa) = (0.011 \pm 0.002)$, or $(1.1 \pm 0.2)\%$, contrast with the microstructural values estimated for fast ramp compressed magnesium at 309, 409 and 563 $GPa$, where $D$ is two to four times lower and $\varepsilon$ is mostly negative with values that are relatively small or even near zero. On the one hand, it could probably reflect the different deformation mechanism active under $P = 959\ GPa$. On the other hand, taking into account a limited number of XRD data and the broad 95% confidence band (Fig. 4), it should be mentioned that the WH analysis at 959 $GPa$ may be less reliable, as more XRD peaks are likely to be needed for a more accurate fit and determination of microstructural parameters.

## 5. Discussion

Previously, the Williamson-Hall analysis was applied to determine the microstrain and crystalline size in statically highly compressed $La_4H_{23}$ [39], $La_3Ni_2O_{7-\delta}$ [40], sulfur [41], $H_3S$ [42], and $BaH_{12}$ [43].

Recently, the Williamson-Hall analysis has been applied to elemental gold and lead subjected to fast ramp compression [22]. The analysis showed that gold at pressure $P = 1003$ GPa, at which it exhibits the *fcc→bcc* phase transition, has large grain sizes (that exceed the resolution of the method) and large microstrain $\varepsilon \sim 0.015$. Meanwhile, elemental lead at $P = 200$ GPa demonstrates the ultrafine structure with an average grain size $D \sim 4$ nm and $\varepsilon \sim 0.006$ that could explain the effect of extreme hardening of pure lead at high pressure and strain rate [14].

In this study, we have made further exploration the application of the WH analysis, and applied it to the XRD data registered for elemental magnesium under fast ramp compression at four pressures – $309\ GPa$, $409\ GPa$, $563\ GPa$ and $959\ GPa$. Following the best practice in data analysis, we need to mention methodological limitations that were used in this study for the WH model. First of all, we were limited in number of available reflections, and, secondly, we used the isotropic WH model because for employing more complicated model more data are required.

Table 1 summarizes the crystalline size, $D$, and microstrain, $\varepsilon$, values which we obtained in this study. Fig. 5 depicts the results in dependence on pressure, $P$, for these characteristics.

**Table 1**. The results of the fits of the $\beta_s(\theta)$ data for fast ramp compressed magnesium at different pressures with the WH equation (Eq. 4) at $D$ and $\varepsilon$ are free fitting parameters and at $\varepsilon$ set to 0 (for $P = 409\ GPa$ and $P = 563\ GPa$).

| pressure, GPa | crystallite size, $D$ (nm) | microstrain, $\varepsilon$ | crystallite size, $D$ (nm), at $\varepsilon = 0$ |
|---|---|---|---|
| 309 | $2.2 \pm 0.7$ | $-0.011 \pm 0.007$ | – |
| 409 | $4.5 \pm 3$ | $-0.003 \pm 0.007$ | $6.1 \pm 0.8$ |
| 563 | $2.6 \pm 0.5$ | $-0.004 \pm 0.004$ | $3.1 \pm 0.1$ |
| 959 | $> 12$ | $0.011 \pm 0.002$ | – |

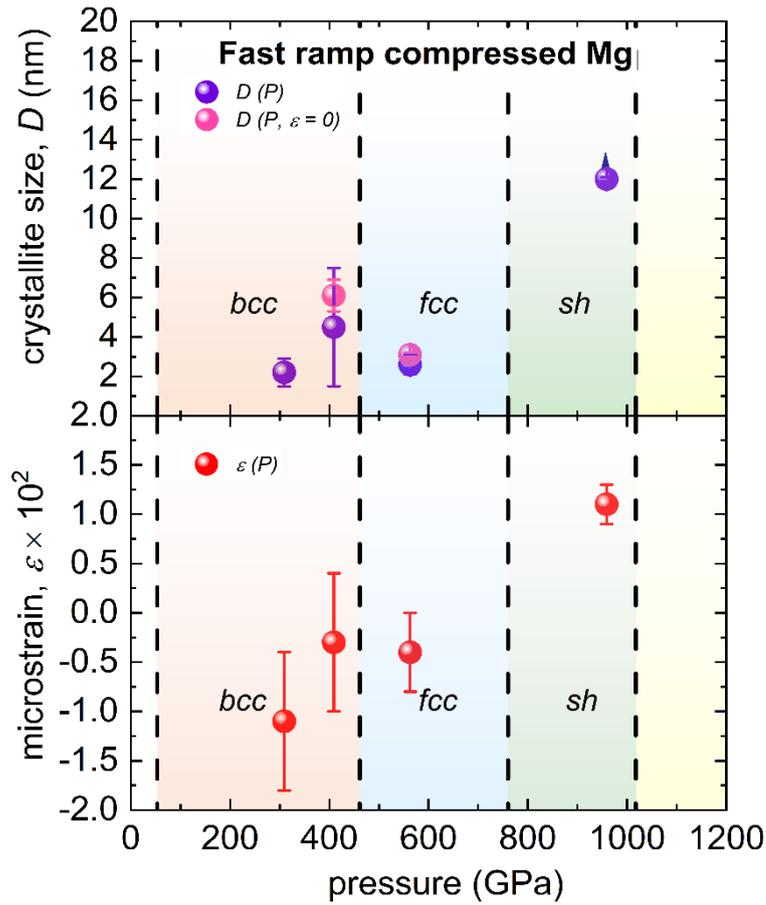

**Figure 5.** Deduced (a) crystalline size $D(P)$ and $D(P, \varepsilon = 0)$, (b) microstrain $\varepsilon(P)$ in fast compressed magnesium.

Particularly at $P = 309\ GPa$, magnesium in *bcc*-like phase exhibits the presence of nanostructure with the crystalline size $D \sim 2$ nm and the microstrain $|\varepsilon| \sim 0.011$. The negative sign of the strain indicates a compressive strain in the magnesium lattice. At $P = 409\ GPa$ and $P = 563\ GPa$, the microstrain errors exceed or equal to the values themselves, justifying constrained fits with $\varepsilon = 0$. These restricted models result in the average crystalline sizes of $\sim 6\ nm$ and $\sim 3\ nm$, respectively, with the absent of lattice strain.

An interesting yield occurs at $P = 959\ GPa$ upon transformation to the *sh*-phase. The crystalline size increases beyond the instrumental resolution limit of $D > 12\ nm$ while microstrain becomes positive with high value of $\varepsilon \sim 1.1\%$. In should be noted that the fast ramp compression occurs on nanosecond timescales and generates significant heating, with temperatures reaching thousands of Kelvin. It could be suggested, under conditions on a nanosecond timescale that may exclude diffusion process, an increase in pressure could stimulate grain growth via a martensitic-like mechanism. In this process, pressure leads to an increase in microstrain due to the forced lattice accommodation. The high density of nuclei formed during the transformation may undergo coalescence where grains joining. This could lead to grain growth and results in the formation of defects at the new phase boundaries.

The grain growth during high-pressure phase transitions has been previously detected in several materials under static compression in diamond anvil cell [44,45], and the phenomenon was attributed to complicated mechanism of combined motion of phase interfaces and grain boundaries under combined thermodynamic driving forces [46]. In those cases, transformation-driven grain growth from several nanometers to micrometers was observed during the transformation process within hours at room temperature. As for fast dynamic compression, it was reported in [47,48] that the grain growth was observed in stishovite under compression over hundred nanosecond time scale.

**Conclusions**

In this study, we reported the first microstructural analysis of elemental magnesium under fast ramp compression using the Williamson–Hall method. Our analysis of recently reported experimental data by Gorman *et al.* [19] showed that at $P = 309\ GPa$ magnesium in *bcc*-like phase has an average grain size of $D = (2.2 \pm 0.7)\ nm$ and microstrain of $\varepsilon = (-0.011 \pm 0.007)$. At $P = 409\ GPa$, *bcc*-like-magnesium demonstrates crystalline size of $D = (4.5 \pm 3)\ nm$ with microstrain of $\varepsilon = (-0.003 \pm 0.007)$. At $P = 563\ GPa$, *Fmmm* phase of magnesium has crystalline size $D = (2.6 \pm 0.5)\ nm$ with microstrain of $\varepsilon = (-0.004 \pm 0.004)$. In case of $P = 959\ GPa$, we revealed that magnesium in *sh* phase exhibits average crystalline size of $D > 12\ nm$ and relatively high value of microstrain of $\varepsilon = (0.011 \pm 0.002)$ in comparison with structural parameters at lower pressures. In the result, we report the first microstructural evolution insights of magnesium under fast ramp conditions.

**Declaration of competing interest**

The authors declare that they have no known competing financial interests or personal relationships that could have appeared to influence the work reported in this paper.

**Acknowledgments**

The work was carried out within the framework of the state assignment of the Ministry of Science and Higher Education of the Russian Federation for the IMP UB RAS. D.A.K. gratefully acknowledged the research funding from the IMP UB RAS junior scientist project No. 17-25. E.F.T acknowledged the research funding from the Ministry of Science and Higher Education of the Russian Federation under Ural Federal University Program of Development within the Priority-2030 Program.

**Author contributions**

A.Yu.V. and E.F.T. conceived the study. D.A.K. performed the analysis, the results of which were confirmed by A.Yu.V. and E.F.T. D.A.K. wrote the manuscript and prepared the figures, which were revised by E.F.T.

**Supplementary Materials**

**Nanocrystalline structure and strain in magnesium under extreme dynamic compression**

**by D.A. Komkova, A.Yu. Volkov, and E.F. Talantsev**

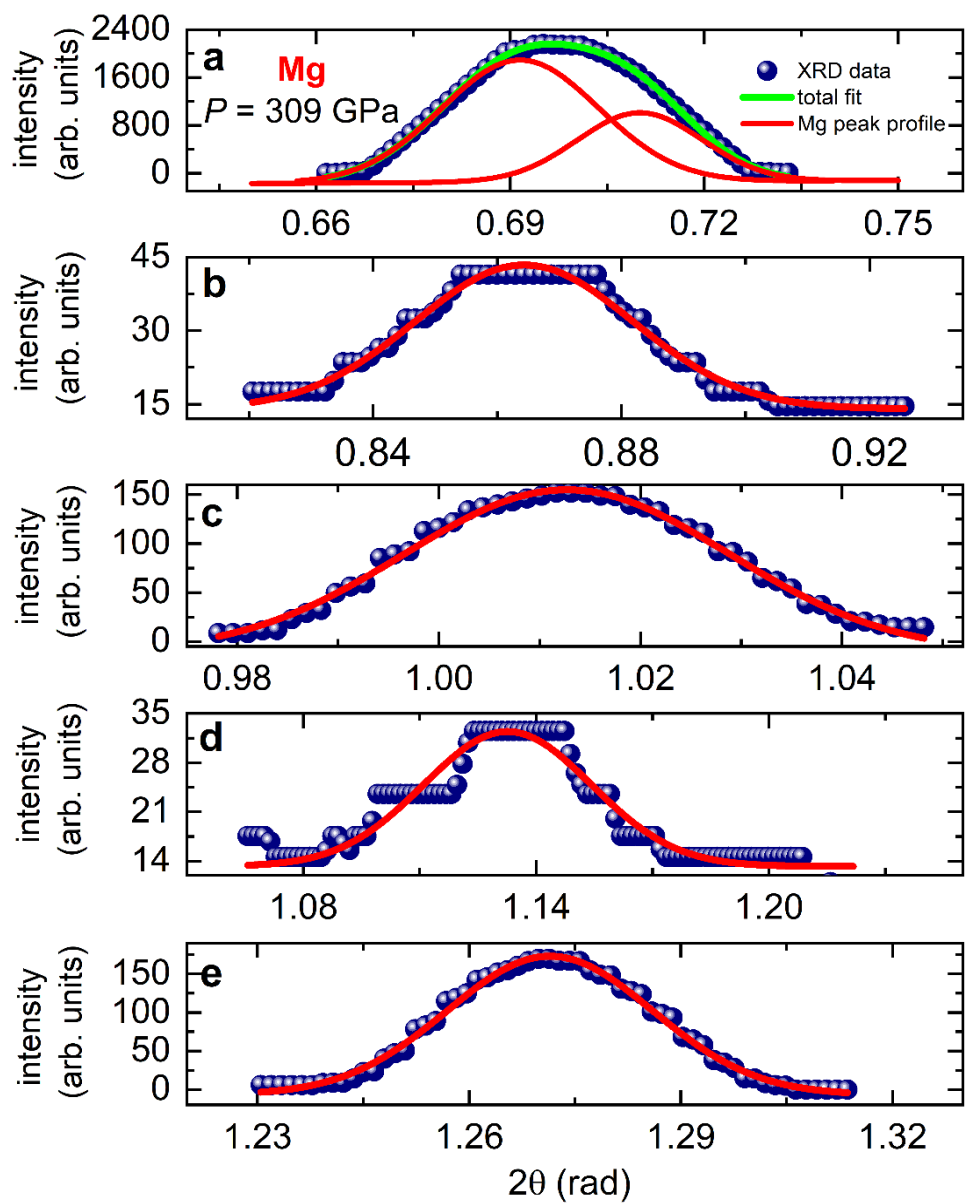

**Figure S1.** XRD data and data fits to the Gauss function (Eq. 1, main text) for peaks of $bcc$-like Mg under fast ramp compression at $P = 309\ GPa$ [19]. Deduced parameters are: (a) $2\theta = (0.6915 \pm 0.0022)\ rad$, $w = (0.0243 \pm 0.0018)\ rad$; $2\theta = (0.7100 \pm 0.0021)\ rad$, $w = (0.0193 \pm 0.0027)\ rad$, fit quality (R-Square (COD)) is 0.9981; (b) $2\theta = (0.8643 \pm 0.0002)\ rad$, $w = (0.0351 \pm 0.0008)\ rad$, fit quality (R-Square (COD)) is 0.9846; (c) $2\theta = (1.0127 \pm 0.0002)\ rad$, $w = (0.0322 \pm 0.0010)\ rad$, fit quality (R-Square (COD)) is 0.9908; (d) $2\theta = (1.1326 \pm 0.0007)\ rad$, $w = (0.0432 \pm 0.0019)\ rad$, fit quality (R-Square (COD)) is 0.9054, (e) $2\theta = (1.2715 \pm 0.0002)\ rad$, $w = (0.0291 \pm 0.0002)\ rad$, fit quality (R-Square (COD)) is 0.9923.

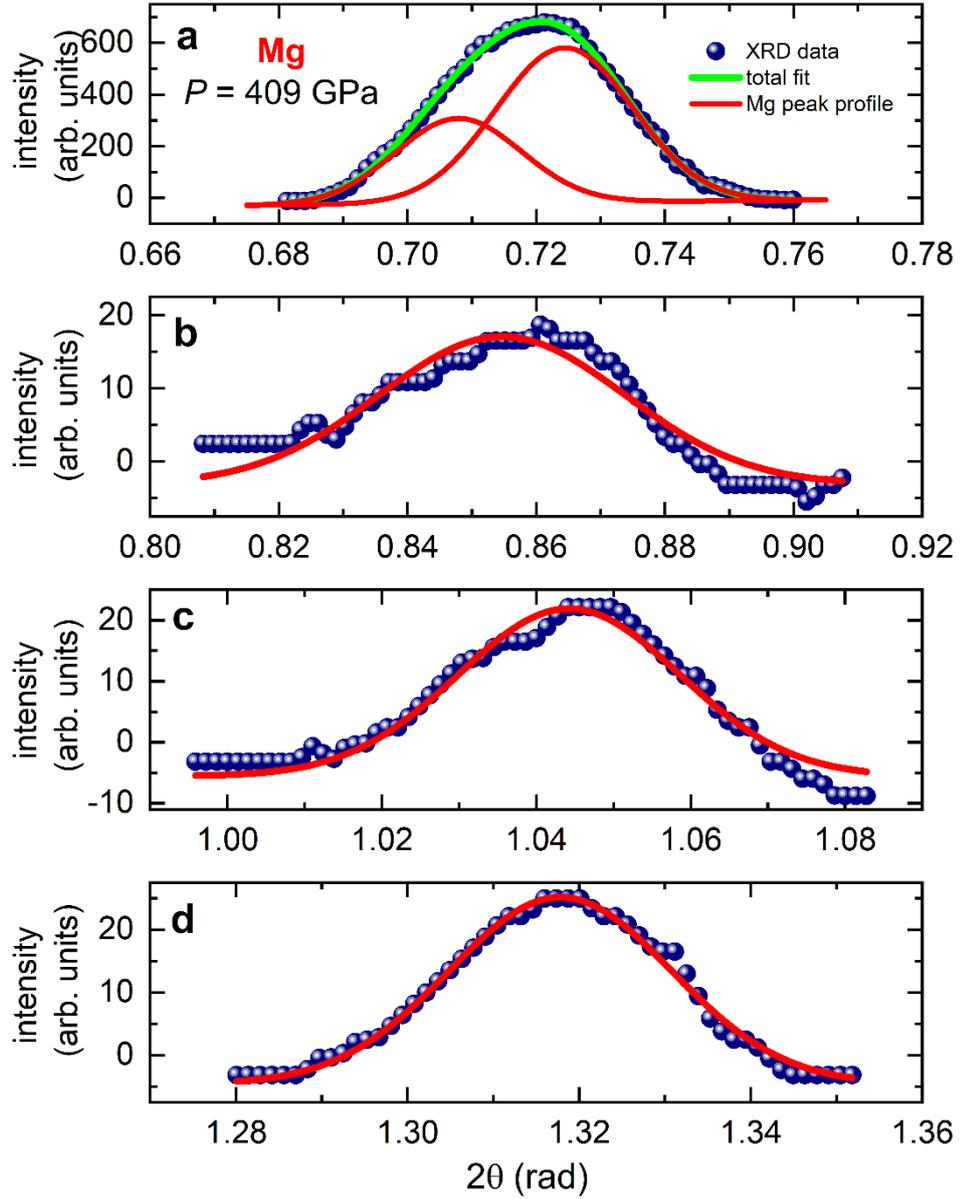

**Figure S2.** XRD data and data fits to the Gauss function (Eq. 1, main text) for peaks of *bcc*-like Mg under fast ramp compression at $P = 409\ GPa$ [19]. Deduced parameters are: (a) $2\theta = (0.7079 \pm 0.0032)\ rad$, $w = (0.0196 \pm 0.0013)\ rad$; $2\theta = (0.7244 \pm 0.0023)\ rad$, $w = (0.0213 \pm 0.0024)\ rad$, fit quality (R-Square (COD)) is 0.9989; (b) $2\theta = (0.8547 \pm 0.0006)\ rad$, $w = (0.0382 \pm 0.0025)\ rad$, fit quality (R-Square (COD)) is 0.9038; (c) $2\theta = (1.0443 \pm 0.0004)\ rad$, $w = (0.0287 \pm 0.0011)\ rad$, fit quality (R-Square (COD)) is 0.9588; (d) $2\theta = (1.3178 \pm 0.0002)\ rad$, $w = (0.0259 \pm 0.0006)\ rad$, fit quality (R-Square (COD)) is 0.9909.

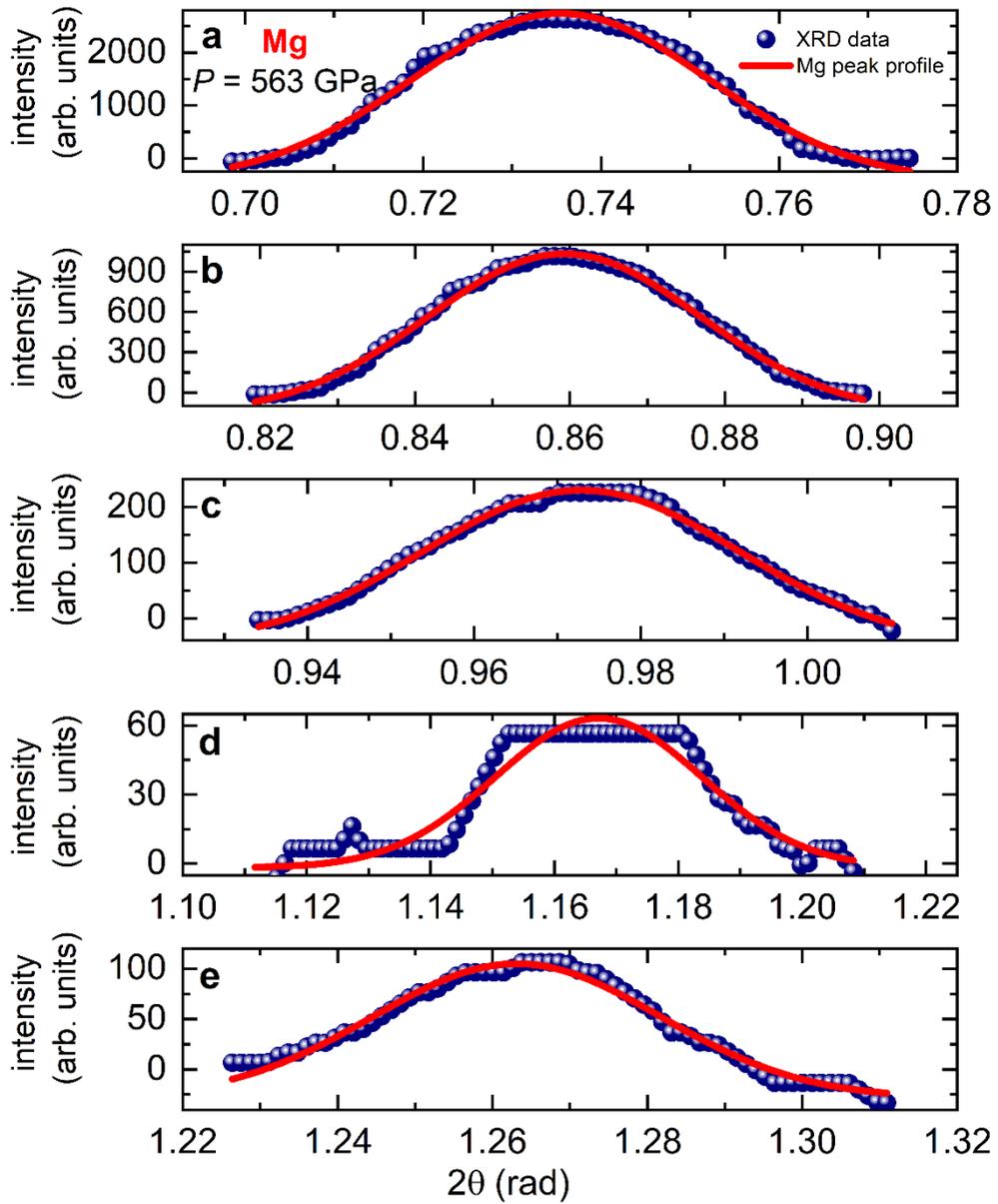

**Figure S3.** XRD data and data fits to the Gauss function (Eq. 1, main text) for peaks of *Fmmm* Mg under fast ramp compression at $P = 563\ GPa$ [19]. Deduced parameters are: (a) $2\theta = (0.7354 \pm 0.0002)\ rad$, $w = (0.0335 \pm 0.0009)\ rad$, fit quality (R-Square (COD)) is 0.9882; (b) $2\theta = (0.8593 \pm 0.0001)\ rad$, $w = (0.0353 \pm 0.0006)\ rad$, fit quality (R-Square (COD)) is 0.9957; (c) $2\theta = (0.9728 \pm 0.0001)\ rad$, $w = (0.0381 \pm 0.0008)\ rad$, fit quality (R-Square (COD)) is 0.9956; (d) $2\theta = (1.1671 \pm 0.0004)\ rad$, $w = (0.0334 \pm 0.0016)\ rad$, fit quality (R-Square (COD)) is 0.9242; (e) $2\theta = (1.2631 \pm 0.0002)\ rad$, $w = (0.0371 \pm 0.0011)\ rad$, fit quality (R-Square (COD)) is 0.9811.

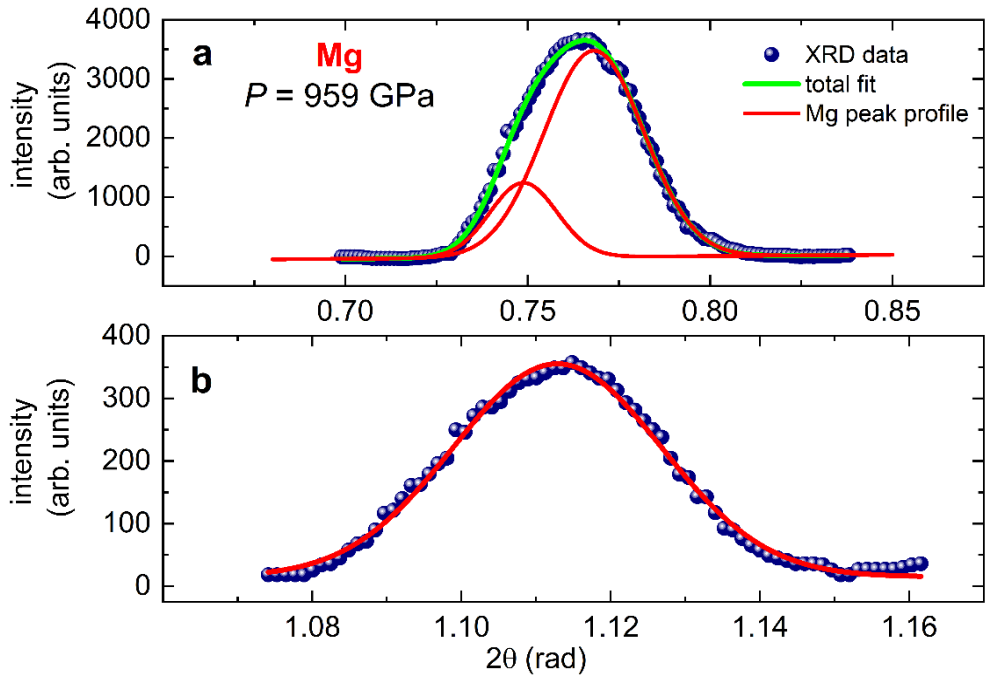

**Figure S4.** XRD data and data fits to the Gauss function (Eq. 2, main text) for peaks of Mg formed under fast ramp compression at $P = 959\ GPa$ [19]. Deduced parameters are: (a) $2\theta = (0.7488 \pm 0.0008)\ rad$, $w = (0.0184 \pm 0.0008)\ rad$; $2\theta = (0.7682 \pm 0.0008)\ rad$, $w = (0.0273 \pm 0.0008)\ rad$, fit quality (R-Square (COD)) is 0.9987; (b) $2\theta = (1.1129 \pm 0.0001)\ rad$, $w = (0.0280 \pm 0.0004)\ rad$, fit quality (R-Square (COD)) is 0.9937.